\documentclass[12pt]{iopart}
\usepackage{iopams}
\usepackage{graphicx}
\begin{document}

\title[Direct calculation of length contraction and clock retardation]{Direct calculation of length contraction and clock retardation}
\author{D V Red\v zi\' c}

\address {Faculty of Physics, University of Belgrade, PO
Box 44, 11000 Beograd, Serbia} \eads{\mailto{redzic@ff.bg.ac.rs}}

\begin{abstract}
For simple electromagnetic models of a rod and a clock, a change of
the shape of the rod and of the rate of the clock when they are set
in uniform motion is calculated exactly, employing the correct
equation of motion of a charged particle in the electromagnetic
field and the universal boostability assumption. Thus it is
demonstrated that, for the simple system considered, length
contraction and clock retardation can be interpreted as dynamical
cause-and-effect phenomena, and not as kinematical effects as is
usually construed in conventional presentations of Special
Relativity. It is argued that the perspectival relativistic change
of an object (corresponding to observations from two inertial
frames), while certainly an acausal effect, has a dynamical content
in the sense that it is {\it tantamount to} an actual dynamical
change of the object in one frame.
\end{abstract}

\section{Introduction}
Recently, I attempted to clarify some basic concepts and results of
Einstein's special relativity theory \cite{DVR1}, noting the
paramount importance of what I called `the universal boostability
assumption' for construction of the theory. The fundamental
assumption reads: `It is possible to set a measuring rod or clock in
a uniform motion or bring it back to a permanent rest without
changing the rest length of the rod or the rest rate of the clock,
i.e. it is possible to boost them in such a way that they remain
standards of length and time in their rest frame, regardless of the
constitution of these objects.' I pointed out that Einstein used a
stronger assumption in his original foundation of Special Relativity
\cite{AE}, namely that the measuring capacity of a measuring rod or
clock remains untouched under {\it arbitrary} boosts; I argued that
the stronger assumption is unwarranted. Particularly, I analysed in
detail the well-known relation for relativistic length contraction,

\begin {equation}
l_v = l_0'\sqrt{1 - v^2/c^2}\, ,
\end {equation}
relating lengths $l_v$ and $l_0'$ of one and the same rod as
measured in two inertial frames $S$ and $S'$ in standard
configuration ($S'$ is uniformly moving at speed $v$ along the
common positive $x, x'$-axes, and the $y$- and $z$-axis of $S$ are
parallel to the $y'$- and $z'$-axis of $S'$), respectively; $S'$ is
the rest frame of the rod, and $S$ is the lab frame, with respect to
which the rod is in uniform motion along its length at speed $v$. I
recalled that Einstein in \cite{AE} stated that if the rod to be
measured is at rest in $S$, then `in accordance with the principle
of relativity' its length as measured in $S$, $l_0$, must be equal
to $l_0'$,

\begin {equation}
l_0 = l_0'\, ,
\end {equation}
employing the same measuring rod as in the earlier measurements.
Equations (1) and (2) imply

\begin {equation}
l_v = l_0\sqrt{1 - v^2/c^2}\, .
\end {equation}
Thus, according to Einstein, a rod initially at rest in an inertial
frame, after a constant velocity $\bi v$ is imparted to it {\it in
an arbitrary way} so that the rod moves freely and uniformly along
its length is contracted (its length is reduced), all with respect
to that frame, as expressed by equation (3). However, I argued in
\cite{DVR1}, noting the relevance of rest properties--preserving
accelerations, that all one may infer on the basis of Special
Relativity is that, in general, equation (1) always applies, whereas
equation (2) and thus equation (3) do not necessarily apply. While
length contraction and clock retardation are generally regarded,
starting from Einstein, as purely kinematical results of Special
Relativity, obtained directly from the Lorentz transformations, I
pointed out that equation (3) (which involves rest
length--preserving accelerations), encapsulates the actual dynamical
change of the rod in the $S$ frame due to the action of some forces
on the rod in that frame. Moreover, even equation (1), which
expresses {\it the relativistic perspectival change} of the rod
(involving measurements from two different frames $S$ and $S'$ and,
clearly, involving no forces acting on the rod by a mere transition
to another inertial frame) has a natural dynamical content.

In a recent paper \cite {DVR2}, I continued my attempts to clarify
Special Relativity. To avoid possible terminological and conceptual
muddle, I proposed to call the contents of equations (1) and (3),
the relativistic length reduction and the relativistic
FitzGerald--Lorentz contraction, respectively. I noted what I
consider to be fallacies in the existing literature devoted to
teaching of relativity, particularly the contention that in the
perspectival change of an object in Special Relativity
(corresponding to observations from two different inertial frames),
there is no change in the object, it is only the reference frame
that is changed from $S$ to $S'$. (More precisely, some authors
explain the differences in observations between two inertial frames
as a purely kinematical effect due to the relativity of
simultaneity, `a consequence of our way of regarding things' (cf,
e.g., \cite{JF,MB}), while other authors (cf, e.g., \cite{EF1,DM})
argue that the differences are basically of a {\it dynamical}
origin, due to a dynamical change of standards of length and time
when transferring the standards between the frames $S$ and $S'$.) On
the one hand, the relativistic perspectival change of an object is
certainly an acausal phenomenon (there is no change in the object in
the standard physicists' sense of the word, referring to different
properties of the object with time in one frame); on the other hand,
as is pointed out in \cite{DVR1,DVR2}, there is a dynamical content
of the phenomenon which seems to be somewhat neglected in the
literature.

The purpose of the present note is to illustrate deliberations
presented in references \cite{DVR1} and \cite{DVR2} with simple
examples, using elementary models of standards of length and time.
Since measuring rods and clocks are physical devices and are subject
to the laws of physics in accordance with which they are
constructed, one must employ physical laws whose validity is well
confirmed in an inertial frame (laboratory). The good candidates are
Maxwell's equations and the Lorentz force expression for the force
acting on a charge $q^*$ in an electromagnetic field,

\begin {equation}
\bi F_L =  q^* \bi E + q^* \bi v \times \bi B\, ,
\end {equation}
where $\bi v$ is the instantaneous velocity of the charge, $\bi E$
is the electric field and $\bi B$ is the magnetic flux density. This
has to be combined with the equation of motion of the charge $q^*$
in the electromagnetic field

\begin {equation}
\frac{d}{dt}\left(\frac {m\bi v}{\sqrt {1-v^2/c^2}}\right) =  q^*
\bi E + q^* \bi v \times \bi B\, ,
\end {equation}
where $m$ is the mass of the charge and $c$ is the speed of light
{\it in vacuo}; the last equation fits the experimental facts if the
{\it additional independent} assumption that $m$ is time-independent
is introduced (cf, e.g. \cite{DVR3}). For simple models of a rod and
a clock, operating on the basis of Maxwell's equations and equation
(5), it will be shown that when the rod and the clock are set in a
uniform motion with respect to the laboratory frame, they exhibit
the FitzGerald--Lorentz contraction and the Larmor clock retardation
in the lab, assuming rest properties--preserving accelerations.
Thus, for the rod and clock under consideration, a dynamical content
of the effects is clearly revealed.

Dynamical analyses of length contraction and clock retardation in
the spirit of the present one, based on electromagnetic laws, have
been published occasionally \cite{JB,OJ1,DM}. Unfortunately, models
proposed in \cite{JB,OJ1,DM} either cannot be solved analytically
\cite{JB}, or introduce for clocks, {\it tacitly} \cite{DM} or
explicitly \cite{OJ1}, a confusing assumption that the velocity of
moving clock is much larger than the maximum velocity occurring in
its clockwork. Thus they are not very convincing. Another point is
that some authors \cite{JB,DM} attempt a constructive dynamical
approach to Special Relativity, what seems to be an impossible
mission, apart from the fact that Maxwell's theory cannot account
for the empirical stability of solid matter. Namely, if one starts
from known and conjectured {\it good} laws of physics in any {\it
one} inertial frame, one can learn that if a constant velocity is
imparted to a rod and a clock, the moving rod is contracted and the
moving clock runs slower. However, rod contraction and clock
retardation are necessary but not sufficient conditions for the
Lorentz transformations: to construct another inertial frame and to
derive the Lorentz transformations, one has to introduce, at one
place or another, Einstein's two postulates of Special Relativity
aided with the universal boostability assumption. (Rod contraction
and clock retardation in the $S$ frame imply that one clock--two way
speed of light is $c$ also in the $S'$ frame but this does not
suffice to `spread time over space' in $S'$.) Moreover, only on the
basis of Einstein's principle approach one knows that candidates for
{\it good} physical laws in one inertial frame must be (or can be
made to be) Lorentz--covariant. (Note that despite repeated
statements in the literature \cite{WR,CM1,EF1}, Lorentz--covariance
of Maxwell's equations is not fulfilled {\it automatically}, as is
pointed out, e.g. in \cite{WGVR,DVR4}.) Thus, it appears that the
laws of physics in any {\it one} reference frame {\it cannot}
`account for all physical phenomena, including the observations of
moving observers,' contrary to Bell's claim in \cite
{JB}.\footnote[1] {Miller in \cite{DM} criticizes Bell's
anticipation of the equation of motion (5) as a limitation of Bell's
approach \cite {JB}. In his own constructive dynamical attempt to
derive Special Relativity, Miller avoided the use of equation (5).
Instead, he tacitly postulated that Maxwell's equations apply not
only in the original rest frame of a physical system, but also in
its final rest frame, cf the argument leading to equation (6) in
\cite{DM}. Thus Miller's approach is at best a combination of a
constructive dynamical, and the principle approaches.}

On the other hand, the standard `kinematical' derivation of rod
contraction and clock retardation \cite{AE}, conceals the
fundamental fact that rods and clocks in hand must be {\it
relativistically valid}, that is, represent physical apparatuses or
devices operating in accordance with the laws of physics which are
(or can be made to be) Lorentz--covariant. In the standard approach,
unexpected qualities of rods and clocks in motion appear as a dry
consequence of the Lorentz transformations, which are achieved from
logically entangled postulates, and which deal with rods and clocks
{\it in abstracto}, regarded as primitive entities (cf \cite{DVR1},
p 199). However, paraphrasing M{\o}ller \cite{CM2}, students of the
theory of relativity would like to see---at least on a simple
model---that rod contraction and clock retardation indeed follow
from the structure of a physical system and the dynamical laws
governing it, considered {\it in one frame only} (cf also
\cite{DVR5}). Therefore I believe that a simple and exact model of a
relativistically valid clock would enhance the students'
understanding of Special Relativity. Also, a dynamical content of
the relativistic perspectival change in an object, and the universal
boostability assumption, seem to be either neglected or
misrepresented in the literature. Thus the present note, which aims
to complement the standard principle approach to Special Relativity
by providing simple illustrations of its dynamical contents, could
perhaps be of some interest.

\section{Direct calculation of length contraction}
Firstly, as a relativistically valid standard of length I discuss an
elementary model of solid body proposed by Sorensen \cite{RS}.
\subsection{Rod at rest}
Consider four equal charges $q$ of the same sign, at rest in the $S$
frame (laboratory), placed at the vertices of a square $ABCD$ of a
side $a$ ($A$ is the bottom left hand vertex, and the vertices $B$,
$C$ and $D$ run counterclockwise). Employing the Coulomb law, one
finds that placing a charge of opposite sign, $q_c = -q(1 +
2\sqrt{2})/4$, at the centre of the square, the resultant of the
forces acting on each charge is zero. Thus the system of the five
charges is in the electrostatic equilibrium. From Earnshaw's
theorem, we know that the equilibrium is unstable (cf, e.g.
\cite{IT}).

One can verify that for five point charges $q,q,q,q$ and $q_c$, the
only static equilibrium shape is a square and not a rectangle or any
other shape, as Sorensen \cite{RS} pointed out. Note that
equilibrium conditions fix only the shape of the equilibrium
configuration and not its size (the side of the square can be
arbitrary). (Incidentally, the electrostatic potential energy of the
static configuration is always zero, regardless of the value of
$a$.)

\subsection{Rod in uniform motion}
Assume now that the system considered has been accelerated, starting
from rest until reaching a steady velocity $\bi v_0 = v_0\hat {\bi
x}$, so that all five charges are uniformly moving in the plane of
the initial square (the $xy$ plane) parallel to the $x$ axis; take
that $\bi v_0$ is perpendicular to the sides $AD$ and $BC$ of the
square. Assume also that that the acceleration was {\it gentle}, in
the sense that, after all transient effects have died out, the
system of five uniformly moving charges is again in a stationary
(time-independent) configuration. The question arises, is there such
a moving configuration at all.

Now we have to take into account that at the location of each
charge, in addition to the electric field, there will be also a
magnetic field, since the remaining charges are in motion. The $\bi
E$ and $\bi B$ fields of a point charge $q$ moving with constant
velocity $\bi v_0$ were first obtained by Oliver Heaviside in 1888
and the $\bi B$ field was rederived by J J Thomson in 1889
(\cite{OH1,OH2,JJT}, cf \cite{OJ2} and references therein), long
before the advent of Special Relativity. The electric field is
radial but not spherically symmetrical (contrary to the
electrostatic field of $q$), and is given by

\begin {equation}
\bi E(\bi r,t) = \frac{q}{4\pi \epsilon_0}\frac {\bi r}{r^3}\frac{1
- v_0^2/c^2}{(1 - v_0^2\sin^2\theta/c^2)^{3/2}}\, ,
\end {equation}
where $\bi r$ is the position vector of a field point with respect
to the instantaneous (at the {\it same} instant t) position of $q$,
$\theta$ is the angle between $\bi r$ and the velocity $\bi v_0$,
and $c^2 \equiv 1/\epsilon_0\mu_0$. (Recall that throughout the
relativity paper \cite{AE}, Einstein used the same symbol ($V$) for
the speed of light {\it in vacuo} and the speed of electromagnetic
waves {\it in vacuo} $(V \equiv 1/\sqrt{\epsilon_0\mu_0})$, linking
thus Special Relativity with Maxwell's theory (cf \cite{DVR1}, p
197).) The magnetic flux density is

\begin {equation}
\bi B(\bi r,t) = \epsilon_0\mu_0 \bi v_0 \times \bi E(\bi r,t)\, .
\end {equation}

Equation (7) and the Lorentz force expression (4) imply that the
total electromagnetic force on each of the five charges of our
uniformly moving system vanishes if and only if the $\bi E$ field
vanishes at the location of each charge. The symmetry suggests that
the equilibrium configuration we are looking for has a rectangular
shape with $q_c$ at the centre. Therefore we assume that stationary
configuration is a uniformly moving rectangle $ABCD$, with sides
$AD$ and $BC$ perpendicular to $\bi v_0$. Denote lengths of the
sides $AB$ (parallel to $\bi v_0$) and $AD$ (transverse to $\bi
v_0$) by $b$ and $d$, respectively. Consider equilibrium conditions
at the vertex $A$. A little analysis reveals that the condition that
the $\bi E$ field at $A$  has no component in the direction
perpendicular to the diagonal $AC$, and along $AC$, implies that

\begin {equation}
\frac{d}{b^2} = \frac {b}{d^2(1 - v_0^2/c^2)^{3/2}}\, ,
\end {equation}
and

\begin {equation}
\frac{1}{b}  + \frac {1}{d(1 - v_0^2/c^2)^{3/2}} = \frac
{2\sqrt{2}}{\sqrt{d^2 + b^2}[1 - v_0^2d^2/c^2(d^2 + b^2)]^{3/2}}\, ,
\end {equation}
respectively. Equation (8) gives

\begin {equation}
b = d\sqrt{1 - v_0^2/c^2}\, .
\end {equation}
It is easy to check that, with this value of $b$, equation (9) is
satisfied identically. As can be seen, equation (10) is necessary
and sufficient condition for the moving rectangular configuration
$ABCD$ to be the equilibrium one, i.e. the stress free state.
(Incidentally, Sorensen {\it assumed} relation (10) from the outset.
Thus he did not demonstrate that `to be in equilibrium [...] the
five charges must have this rectangular shape, shortened in the
dimension of the direction of motion by the Lorentz contraction as
compared to the transverse direction,' contrary to his claim in
\cite{RS}. Instead, he proved only that equation (10) is sufficient
condition for the moving rectangle to be in equilibrium.)

Note that the character of forces governing the equilibrium is such
that equilibrium conditions determine the shape of the configuration
and not its size ($d$ is arbitrary), analogously to the
electrostatic case. Thus, accelerating square of side $a$ until
reaching the steady velocity $\bi v_0$, one can arrive at a moving
stationary rectangle with sides $a\sqrt{1 - v_0^2/c^2}$ and $a$ in
the direction of motion and transverse to it, respectively, but also
with sides  $d\sqrt{1 - v_0^2/c^2}$ and $d$, where $d\neq a$.
Clearly, only in the first case acceleration was rest
length--preserving. Namely, according to Special Relativity,
observing the moving rectangles from their rest frame $S'$, they
will be squares of sides $a$ and $d\neq a$, respectively, since
Maxwell's equations can be made to be Lorentz--covariant, and the
Lorentz force expression should apply in every frame. Thus equation
(1) always applies, whereas equations (2) and (3) do not necessarily
apply, as is pointed out in \cite{DVR1}.

As a historical aside, recall that Lorentz argued long ago that if
to a system $\Sigma'$ of particles in the equilibrium configuration
at rest relative to the ether `the velocity $\bi v = v\hat {\bi x}$
is imparted, it will {\it of itself} change into the system
$\Sigma$' which is got from $\Sigma'$ by the deformation $(\sqrt {1
- v^2/c^2},1,1)$ (\cite{POR}, pp 5--7, 21--23, 27--28, cf also
\cite{DVR2}, pp 60--1). However, Lorentz was wrong here; the change
$\Sigma' \rightarrow \Sigma$ can also be effectuated by the
transformation $(l\sqrt {1 - v^2/c^2},l)$, where $l \neq 1$, as the
present Section 2 reveals. (From $vdl/dv = 0$, Lorentz deduced that
$dl/dv = 0$, $l = const$, and concluded: `The value of the constant
must be unity, because we know already that, for $v = 0$, $l = 1$'
(\cite{POR}, p 27). But, all one can deduce from $vdl/dv = 0$ is
that $dl/dv = 0$ for $v \neq 0$! Thus $l$ may have arbitrary
(constant) value for $v \neq 0$.)

\section{Direct calculation of clock retardation}
The same equilibrium system of five charges, providing a standard of
length in the preceding section, will be employed as an exact and
yet simple model of a relativistically valid clock.

\subsection{Clock at rest}
Let four identical charges $q$ be now fixed at the vertices of the
square $ABCD$ of side $a$ at rest in the lab frame $S$. Denote the
axis perpendicular to the plane of the square which passes through
its centre as the $z$ axis; choose the origin at the centre  and the
$x$ and $y$ axes parallel to the sides $AB$ and $AD$ of the square,
respectively. Remove the charge $q_c$ from its central equilibrium
position to the point on the positive $z$ axis with $z = \cal A$ and
release it with zero initial velocity to move under the action of
the electrostatic field of the remaining four charges.

The exact equation of motion of the charge $q_c$ in the
electrostatic field is

\begin {equation}
m\frac{d}{dt}\left(\frac {\bi v}{\sqrt {1-v^2/c^2}}\right) =  q_c
\bi E \, ,
\end {equation}
where the mass $m$ of $q_c$ is assumed to be time-independent, as is
pointed out in the Introduction. Equation (11) and identity

\begin {equation}
\bi v \cdot\frac{d}{dt}\left(\frac {\bi v}{\sqrt {1-v^2/c^2}}\right)
\equiv c^2\frac{d}{dt}\left(\frac {1}{\sqrt {1-v^2/c^2}}\right)\, ,
\end {equation}
imply that

\begin {equation}
mc^2\frac{d}{dt}\left(\frac {1}{\sqrt {1-v^2/c^2}}\right) = q_c\bi
E\cdot \bi v\, .
\end {equation}
Specifying to our problem, $\bi E$ is the electrostatic field given
by

\begin {equation}
\bi E(0,0,z) = \frac{\kappa 4qz \hat {\bi z}}{(z^2 + a^2/2)^{3/2}}\,
,
\end {equation}
where $\kappa = 1/4\pi\epsilon_0$, and $\bi v = v_z\hat {\bi z}$,
since the motion is along the $z$ axis. Using equations (11), (13)
and (14) one obtains\footnote[2] {The force $q_c\bi E$ is always
parallel to the instantaneous velocity $\bi v$ of the charge $q_c$
so that one can derive equation (15) using the concept of
`longitudinal' mass, taking into account that the Lorentz force
expression is a {\it pure} force (cf \cite{WR,DVR3}). I preferred
not to employ here the obsolete concepts of `transverse' and
`longitudinal' mass, while they appear occasionally in the
literature \cite{WR,OJ1}.}

\begin {equation}
\frac {dv_z}{dt} = \frac {q_c}{m}\left ( 1 - \frac
{v_z^2}{c^2}\right )^{3/2}\frac {\kappa 4qz }{(z^2 + a^2/2)^{3/2}}\,
.
\end {equation}
Obviously, the charge $q_c$ does not perform a simple harmonic
motion. However, noting that $q_cq < 0$, and also that at $t = 0$,
$z = \cal A$, $v_z = 0$, a little analysis reveals that the solution
of equation (15) satisfying these initial conditions is a periodic
function of $t$,

\begin {equation}
z_R = f(t)\, ;
\end {equation}
subscript $R$ serves as a reminder that the square is at rest.
Denote the period of that function by $T_0$; the period comprises
continuous changes of position of the charge $q_c$ from $z = \cal A$
to $z = -\cal A$ and {\it vice versa}. Clearly, the square and the
charge may be considered as a simple model of a clock, and can be
used for measuring time in terms of the number of periods $T_0$.
Namely, as Jefimenko pointed out \cite{OJ1}, `as a physical entity,
time is defined in terms of specific measurement procedures, which
for the purpose of the present discussion may be described simply as
``observing the rate of the clocks.''' However, for observing the
rate of a clock in uniform motion with respect to the lab frame $S$,
one has first to `spread time over space,' i.e. to synchronise
distant clocks at rest in $S$. From Maxwell's equations one knows
that a natural means for that purpose is the propagation of
electromagnetic waves {\it in vacuo}.

\subsection{Clock in uniform motion}
Assume now that the same clock is set in uniform motion with
constant velocity $\bi v_0 = v_0 \hat {\bi x}$ along the positive
$x$ axis, so as to be {\it relativistically} valid, i.e. to serve as
an {\it identical} standard of time also for a co-moving inertial
observer. From the preceding discussion it follows that now four
identical charges $q$  have to be fixed in their rest
length--preserving {\it equilibrium} positions, that is at the
vertices of the moving rectangle $ABCD$ with sides $AB = a \sqrt {1
- v_0^2/c^2}$ and $AD = a$. Remove the charge $q_c$ co-moving with
the rectangle from its central equilibrium position, to the
co-moving point on the axis of the rectangle with $z = \cal A$, and
release it with initial velocity $\bi v_0$ to move under the action
of the electromagnetic field of the remaining four charges.

The exact equation of motion of the charge $q_c$ in the field,
obtained from equation (5) assuming that $m$ is constant, reads

\begin {equation}
m\frac{d}{dt}\left(\frac {\bi v}{\sqrt {1-v^2/c^2}}\right) =  q_c
\bi E + q_c \bi v \times \bi B\, ;
\end {equation}
obviously, equation (13) applies in this case too. Using equation
(6), after a somewhat cumbersome but in every step simple
calculation, one finds that the electric field on the co-moving axis
(which is perpendicular to the plane of the moving rectangle $ABCD$
and passes through its centre) is

\begin {equation}
\bi E = \frac{\kappa 4qz \hat {\bi z}}{(z^2 + a^2/2)^{3/2}}\frac
{1}{\sqrt {1-v_0^2/c^2}} = E_z \hat {\bi z}\, ,
\end {equation}
For the magnetic flux density at the same point on the co-moving
axis, using equation (7), one finds

\begin {equation}
\bi B = - \frac{v_0}{c^2}E_z \hat {\bi y} = -
\frac{v_0}{c^2}\frac{\kappa 4qz \hat {\bi y}}{(z^2 +
a^2/2)^{3/2}}\frac {1}{\sqrt {1-v_0^2/c^2}} = B_y \hat {\bi y}\, .
\end {equation}

Using equations (17), (13), (18) and (19), and taking into account
initial conditions (at the time $t = 0$, the charge $q_c$  is at the
point $z = \cal A$ on the co-moving axis, and components of its
velocity are $v_x = v_0$, $v_y = v_z = 0$), a little analysis
reveals that the charge  $q_c$ will forever move along the co-moving
axis, i.e. so that $v_x = v_0$, $v_y = 0$ . Passing details, we give
the final equation of motion of the charge along the co-moving axis:

\begin {equation}
\frac {dv_z}{dt} = \frac {q_c}{m}\left ( 1 - \frac {v^2}{c^2}\right
)^{3/2}\frac {\kappa 4qz }{(z^2 + a^2/2)^{3/2}}\frac {1}{\sqrt
{1-v_0^2/c^2}}\, .
\end {equation}
Now since $v^2 = v_0^2 + v_z^2$ one has

\begin {equation}
\left ( 1 - \frac {v^2}{c^2}\right )^{3/2} = \left ( 1 - \frac
{v_0^2}{c^2}\right )^{3/2}\left [ 1 - \frac {v_z^2}{c^2(1 -
v_0^2/c^2)}\right ]^{3/2}\, ,
\end {equation}
and introducing

\begin {equation}
v_z^* = \frac {v_z}{\sqrt {1-v_0^2/c^2}} = \frac {dz}{dt^*}\, .
\end {equation}
where

\begin {equation}
t^* \equiv t\sqrt {1-v_0^2/c^2}\, .
\end {equation}
equation (20) can be recast as

\begin {equation}
\frac {dv_z^*}{dt^*} = \frac {q_c}{m}\left ( 1 - \frac
{v_z^{*2}}{c^2}\right )^{3/2}\frac {\kappa 4qz }{(z^2 +
a^2/2)^{3/2}}\, .
\end {equation}

Equation (24) for the clock in motion has exactly the same form as
equation (15) for the {\it same} clock at rest, the only difference
being that variable $t$ in the latter is replaced by $t^* \equiv
t\sqrt {1-v_0^2/c^2}$ in the former. Since the solution of equation
(15) is a periodic function (16)  with period $T_0$, it is clear
that the solution of equation (24) satisfying identical initial
conditions ($z = \cal A$ and $v_z = v_z^* = 0$, at $t = t^* = 0$) is
the same function of $t^*$,

\begin {equation}
z_M = f(t^*) = f(t\sqrt {1-v_0^2/c^2})\, ,
\end {equation}
where subscript $M$ serves as a reminder that the clock is in
uniform motion; period of the clock in motion is obviously

\begin {equation}
T_M = \frac {T_0}{\sqrt {1-v_0^2/c^2}}\, .
\end {equation}
Thus, the above model of a clock, however fragile, provides a simple
and yet exact illustration of the Larmor clock
retardation.\footnote[3] {Note that when the clock (the square + the
charge $q_c$) moves {\it along} the axis of the square (the $z$
axis),  so that $\bi v_0 = v_0 \hat {\bi z}$ (in the same way as
Jefimenko's clock $\#$ 1 \cite{OJ1}), an {\it exact one frame}
derivation of equation (26) appears rather challenging.} For this
clock, the retardation appears to be a dynamical, cause--and--effect
phenomenon, as was the case with length contraction discussed in
Section 2.\footnote[4] {What is the origin of labelling rod
contraction and clock retardation as kinematical effects (i.e., that
they can be dealt with without involving actions, forces, masses)?
This appears to be relics of Newton's absolute space and time
concepts, where it is tacitly assumed that `a moving rigid body at
the epoch $t$ may in geometrical respects be perfectly represented
by the {\it same} body {\it at rest} in a definite position'
\cite{AE}, and analogously for a moving clock. While `kinematical'
is fitting in the context of the Galilean transformation, it
masquerades the dynamical contents of the Lorentz transformation.} A
more detailed analysis of clock retardation is given in next
Subsection.

Rewrite equation (26) in the form

\begin {equation}
T_v = \frac {T_0}{\sqrt {1-v^2/c^2}}\, ,
\end {equation}
where now $v$ is the speed of the rectangle in uniform motion, so as
to be analogous to equation (3). Since rest rate--preserving
acceleration is assumed in the above derivation of equation (27),
one has that

\begin {equation}
T_0 = T_0'\, ,
\end {equation}
where $T_0'$ is period of the moving clock as observed by an
inertial co-moving observer. Equations (27) and (28) imply

\begin {equation}
T_v = \frac {T_0'}{\sqrt {1-v^2/c^2}}\, ,
\end {equation}
which is analogous to equation (1). As can be seen, {\it mutatis
mutandis}, remarks analogous to those for the uniformly moving rod,
presented in the last two paragraphs of Section 2, apply to the case
of the uniformly moving clock. Particularly, equation (29) always
applies, whereas equations (28) and (27) need not apply.

Note that the above dynamical derivation of equation (26) applies to
the simple clock considered. On the other hand, in the framework of
relativistic {\it kinematics}, M{\o}ller argued: `In view of the
fact that an arbitrary physical system can be used as a clock, we
see that any physical system which is moving relative to a system of
inertia must have a slower course of development than the same
system at rest' \cite{CM1}. Here one has to take into account that,
according to Special Relativity, any physical system must conform to
some Lorentz--covariant {\it dynamical} laws, however complex the
system is. Since the exact form of the laws is generally unknown,
`an all-inclusive dynamic (causal) interpretation of time dilation
is hardly possible,' as Jefimenko pointed out \cite{OJ1}.
Fortunately, the principle approach to Special Relativity predicts
equation (26) {\it indirectly}, via the Lorentz transformations,
without the need to enter into details of the phenomenon that serves
as a clock. Namely, one need not know the exact laws  governing the
operation of a clock; it suffices to know that the laws have to be
Lorentz--covariant. However, one must admit that any clock
retardation hides a complex dynamical process and also involves the
universal boostability assumption. Finally, note that the above
simple clock model illustrating a dynamical content of equation (26)
represents an {\it ideal} clock. Namely, a real clock necessarily
involves damping, which is in our case due to the radiation reaction
force. Construction of Special Relativity requires of course ideal
standard clocks so our simple model may perhaps be to the point.

\subsection{Clock retardation in detail}
Equation (15) can obviously be recast into
\begin{equation}
\frac{dv_z}{dz}v_z = - \frac{|q_cq|\kappa 4}{m}\left (1 -
\frac{v_z^2}{c^2}\right )^{3/2}\frac{z}{(z^2 + a^2)^{3/2}}\, .
\end{equation}
Separating variables and integrating, setting $v_z = 0$ when $z =
\pm \cal A$ and solving for $v_z$ yields
\begin{equation}
v_z = \frac{dz}{dt} = \mp c \{...\}^{1/2}\, ,
\end{equation}
where
\begin{equation}
\fl \{...\} \equiv \{1 - [1+ (|q_cq|\kappa 4/mc^2)(1/\sqrt {z^2 +
a^2/2} - 1/\sqrt {{\cal A}^2 + a^2/2} )]^{-2} \} \, ,
\end{equation}
and $-$ and $+$ sign corresponds to the motion of $q_c$ in the
direction of decreasing $z$ and increasing $z$, respectively.
Equation (31) implies that, for the oscillator at rest, passage of
$q_c$ from $z$ to $z + dz$ lasts time interval
\begin{equation}
dt = \mp (1/c)dz \{...\}^{-1/2}\, .
\end{equation}
Thus, the period $T_0$ of the oscillator at rest is given by

\begin{equation}
T_0 = \frac{2}{c} \int_{-\cal A}^{\cal A} \{... \}^{-1/2}\rmd z\, .
\end{equation}
Incidentally, in the case of {\it small} oscillations, i.e. when
${\cal A} \ll a$, from equation (34) one obtains that

\begin{equation}
T_0 \approx \frac{2}{c} \int_{-\cal A}^{\cal A} \{1 - [1+ ({\cal
K}/2mc^2)({\cal A}^2 - z^2)]^{-2} \}^{-1/2}\rmd z\, .
\end{equation}
where ${\cal K} \equiv |q_cq|\kappa 4/(a^2/2)^{3/2}$. As can be
seen, this result coincides with the corresponding M{\o}ller's
result for the period of his ```relativistic'' oscillator', equation
(60) in \cite{CM2}, as it should. Finally, note that when ${\cal K}
{\cal A}^2 \ll mc^2$, from equation (35) one obtains the familiar
expression for the period of the simple harmonic oscillator, $T_0 =
2\pi \sqrt {m/{\cal K}}$.

Analogously, using equations (22)-(24), for the oscillator in
uniform motion one finds that passage of $q_c$ from $z$ to $z + dz$
lasts time interval
\begin{equation}
dt_M = \frac{1}{\sqrt {1 - v_0^2/c^2}}(\mp) (1/c)dz \{...\}^{-1/2}\,
,
\end{equation}
which is $1/\sqrt {1 - v_0^2/c^2}$ times longer than the
corresponding time interval (33) for the oscillator at rest. This
embodies clock retardation. The period of the oscillator in motion
is obviously given by
\begin{equation}
T_M = \frac{1}{\sqrt {1 - v_0^2/c^2}}\frac{2}{c} \int_{-\cal
A}^{\cal A} \{... \}^{-1/2}\rmd z\, ,
\end{equation}
which is equation (26).

\section{Where do the perspectival changes come from?}
To avoid confusion, begin with a few terminological comments.

By `the perspectival relativistic change' of an object I mean that,
according to Special Relativity, one and the same object (in the
sense consisting of the same `atoms') has distinct properties (say,
the length of a rod or the period of a clock), depending on whether
the properties are being measured in the rest frame of the object or
in the laboratory frame (with respect to which the object is in a
uniform translatory motion). Note that here we deal with two
different states of motion of an object with respect to two inertial
frames (`observers'), and two respective `configurational states' of
the object. By `the actual physical change' of an object in the
standard physicists' sense of the phrase, I mean that properties of
the object become distinctly different under the action of certain
forces (both external and internal), all with respect to one and the
same inertial frame.

Now comes what is perhaps the key question of Special Relativity.
When a single physical object is observed by two different inertial
observers (or when a single observer changes frames), where do the
differences between their observations come from? Particularly, when
an object at rest in $S'$ and thus in uniform motion with respect to
the lab frame $S$ is observed from the two frames, why the results
of observations differ? Note that the perspectival change itself has
nothing to do with {\it previous} history of the object either in
$S'$ or in $S$, the history may be unknown to us; moreover, the
object need not be free nor connected. Note also that there seems to
be an overall consensus that we did nothing to the object by merely
observing it from two different frames (or by accelerating an
observer to another frame). Thus, there is no cause of the
perspectival relativistic change, it is an acausal effect.

Miller argued in \cite{DM} that the differences among observations
of different inertial observers `are due to the differences in their
respective measuring instruments [...] these perspectival effects
ultimately have a dynamical origin because the properties of
measuring instruments are determined by the forces that keep them in
equilibrium in their respective frames.' The author explained,
following Fe\u{\i}nberg \cite{EF1}, that `when the measuring rods
and clocks are moved between inertial observers, they suffer
dynamical changes. When the observers use their dynamically altered
rods and clocks to make measurements, it is not surprising that
their results differ and that they differ by the same factors that
are involved in the dynamical changes.'

While Fe\u{\i}nberg  and Miller advocate a force interpretation of
the so-called kinematical effects of Special Relativity, a common
thread in their discussions is that `there are no dynamical effects
in the physical object being observed'; the differences in measuring
instruments used by by different inertial observers are all that
matters. Now, it is certainly true that nothing at all has happened
to the object being observed by a mere transition to another
inertial frame. (`The body received no impact, pull or boosts, but
is viewed from a system moving relative to it; [...] there has been
no actual change in the body itself.') However, I think that
Fe\u{\i}nberg  and Miller's interpretation falsifies the spirit of
Special Relativity. While the perspectival relativistic change is an
acausal effect, I will argue that there is a dynamical explanation
of the effect in the sense that the perspectival change is {\it
tantamount to} an actual physical change. This seems to be the gist
of Special Relativity.

Firstly, each inertial observer possesses his or her own set of
measuring instruments which are identical to one another. A
measuring rod at rest in the lab frame $S$ is in all respects
identical to a measuring rod of the same construction at rest in the
`moving' frame $S'$ under identical physical conditions; the rods
{\it embody} the same length in their respective rest frames. That
the rods can indeed be {\it of the same construction} is secured by
the universal boostability assumption, as was illustrated in Section
2.\footnote[5] {According to conventional presentations of Special
Relativity, measuring instruments need not be transferred between
frames; instead, they can be constructed in each frame `from
scratch', following the same recipe. But it appears that the
universal boostability assumption, or its equivalent, must be
introduced at some step of the procedure. At first sight, the
universal boostability assumption has basically the same contents as
Born's `principle of the physical identity of the units of measure'
\cite{MB} (cf also \cite{DVR1}, footnote 12). However, Born seems to
to imply that measuring capacity of a rod or clock remains untouched
under {\it arbitrary} boosts, which is incorrect, cf Section 2. In
the same way, Fe\u{\i}nberg's `{\it universality with respect to the
acceleration regime} [of the rod contraction and the clock
retardation]' \cite{EF2} does not generally hold. } Therefore, it is
somewhat perplexing to account for the differences between the
observations of the $S$-- and $S'$--observer in terms of the
differences in their respective measuring instruments, as
Fe\u{\i}nberg \cite{EF1} and Miller \cite{DM} do. A natural
explanation appears to be at hand.

As was noted above, in the perspectival relativistic change we deal
with two different states of motion of an object with respect to two
inertial observers, and two respective `configurational states' of
the object. Specifying to the simple system discussed in Sections 2
and 3, taking into account that the theory employed (Maxwell's
equations plus the Lorentz force equation (5)) is made to be
Lorentz--covariant, it follows that different observations of the
system considered in the lab frame $S$ and in the rest frame $S'$
are due to different states of motion of the system in the two
frames, and thus to its different respective configurations. The
differences in configurations are due to a different electromagnetic
field produced by the moving field--producing charges, and hence to
a different force acting on the moving field--experiencing charges.

On the other hand, if we start from the object at rest in the lab
frame $S$, which is in the same configurational state as it was the
object's rest state in the `moving' frame $S'$, and accelerate it
until reaching the steady velocity $\bi v = v\hat {\bi x}$ in a
persistent state (thus being at rest with respect to the `moving'
frame) {\it and} if the acceleration was rest
properties--preserving, we reach the same configurational state of
the moving object as measured in the lab, as was found earlier as a
result of the perspectival change. Now we deal with two
configurational states of the same object, which are identical to
the ones discussed above in the context of the perspectival change,
corresponding to two different states of motion but now with respect
to one inertial frame. In the `one frame scenario', one has two {\it
stationary} configurational states with distinct properties of the
object due to different character of forces providing equilibrium in
the two states of motion; this actual physical change has a clear
dynamical origin. Since the `two frames situation' is perfectly
equivalent to the corresponding `one frame situation' under the
assumptions stated, one must admit that not only the actual physical
change but also the perspectival relativistic change of an object
has a dynamical content. (Clearly, in the system discussed in
Section 2, the change belongs to {\it statics} (equilibrium of
forces and momenta).)

Basically, all that matters is the state of uniform motion of a
physical system with respect to {\it one} (arbitrary but fixed)
inertial observer, under the proviso that the (persistent) rest
configuration of the system is conserved. It is irrelevant whether
two different states of motion of the system are observed from two
different inertial frames, respectively, or from one frame only, if
in the latter case the two states of motion are related by a rest
properties--preserving acceleration. This supremacy of any one
inertial observer appears to be the gist of the principle of special
relativity.\footnote[6] {Incidentally, Fe\u{\i}nberg \cite{EF1}
pointed out that, instead of transferring a physical object from its
initial rest frame $S$ to its final rest frame $S'$ through a rest
properties--preserving acceleration, the same final state of the
object in motion relative to $S$ can be reached in a different way.
Namely, instead of accelerating the object, one can accelerate  a
reference frame--copy of the $S'$ frame, initially at rest relative
to $S'$, until the copy frame eventually becomes the $S$ frame (cf
\cite{AE} and also a somewhat obscure attempt by Swann
\cite{SWANN}); of course, the object is now assumed to be always at
rest in $S'$. Fe\u{\i}nberg asked why does the action on the
measuring system of rods and clocks cause a contraction of the
measured rod. After an explanation that I found obscure, he noted
that `one may naturally still wonder why a symmetric result is
obtained when there is such an enormous asymmetry in the transition
to the final state of motion with the same relative velocity.' It
seems, however, that in his explanation Fe\u{\i}nberg failed to take
into account properly that the perspectival change is an acausal
phenomenon, and also that measuring capacity of a rod or clock
remains untouched under rest properties--preserving boosts.}

\section {Summary}
The calculations of rod contraction and clock retardation presented
in this paper provide a dynamical cause--and--effect type
interpretation of those so-called kinematical effects of Special
Relativity. A dynamical content of the effects is clearly revealed
at least in the case of the simple electromagnetic model employed,
in terms of various character of forces governing the equilibrium in
the state of motion and in the state of rest of the system under
consideration. By means of the same model, the importance of the
universal boostability assumption is illustrated. A dynamical
content of the perspectival relativistic change is also discussed.
It is argued that when a connected physical object in a persistent
state is observed by observers in different inertial frames, the
differences among their observations are due to changes in character
of forces which determine the structure of the object with a change
of its velocity, provided that the velocity change was performed
{\it in a rest properties-preserving way}. The different inertial
observers have a dynamical explanation of the differences among
their observations in terms of an equivalent dynamical change in the
object with respect to one inertial frame.

\section*{Acknowledgments}
I thank Vladimir Hnizdo and Alberto Mart\'{i}nez for illuminating
comments on an earlier draft. I also thank Robert Shuler Jr. for
bringing reference [25] to my attention. My work is supported by the
Ministry of Science and Education of the Republic of Serbia, project
No. 171028.

\end{document}